\documentclass[11pt,letterpaper]{JHEP3}

\usepackage{amsmath,amsthm,amsfonts,amscd}
\usepackage{amssymb}
\usepackage{subfigure}
\usepackage{psfrag}
\usepackage{epsfig}
\usepackage{graphics}

\usepackage[latin1]{inputenc}
\usepackage{graphicx}

\DeclareSymbolFont{AMSa}{U}{msa}{m}{n}
\DeclareSymbolFont{AMSb}{U}{msb}{m}{n}
\let\Box\relax
\DeclareMathSymbol{\Box}{\mathord}{AMSa}{"03}

\def\unit{\relax{\rm 1\kern-.26em I}}

\def \eqn#1#2{\begin{equation}#2\label{#1}\end{equation}}

\title{Meta-Stable Supersymmetry Breaking in a Cooling Universe}
\author{Willy Fischler$^\dag$, Vadim Kaplunovsky$^\dag$, Chethan 
Krishnan$^\ddag$, Lorenzo
Mannelli$^\dag$, Marcus Torres$^\dag$\\
$^\dag$Theory Group, Department of Physics \\ University of Texas, Austin,
TX 78712 \\
E-mail: \email{fischler, vadim, lorenzo, \\
\ \ \ \ \ \ mctorres@physics.utexas.edu}\\
\\
$^\ddag$International Solvay Institutes,\\
Physique Th\'eorique et Math\'ematique,\\
Universit\'e Libre
de Bruxelles, \\ C.P. 231, B-1050, Bruxelles, Belgium\\
E-mail: \email{ckrishna@ulb.ac.be}
}

\abstract{We look at the recently proposed idea that susy breaking can
be accomplished in a meta-stable vacuum. In the context of one of the
simplest models (the Seiberg-dual of super-QCD), we address the following
question: if we look at this theory as it cools from high temperature, is
it at all possible that we can end up in a
susy-breaking meta-stable vacuum? To get an idea about the answer, we look at the free energy of the system at
high temperature. We conclude that the phase-structure of the free-energy
as the temperature drops, is indeed such that there is a second order
phase transition in the direction of the non-susy vacuum at a
finite $T=T_c^Q$. On the other hand, the potential barrier in
the direction of the susy vacuum is there all the way till $T \sim
0$.}


\received{???????? ?st, 2000} \accepted{???????? ?th, 1998}
\preprint{hep-th/0611018\\UTTG-15-06\\ULB-TH/07-08}

\begin{document}

%%%%%%%%%%%%%%%%%%%%%%%%%%%%%%%%%%%%%%%%%%%%%%%%%%%%%%%%%%%%%%%%%%%%%%%%%%%%
%          Table of contents automatic !!!                                 %

%%%%%%%%%%%%%%%%%%%%%%%%%%%%%%%%%%%%%%%%%%%%%%%%%%%%%%%%%%%%%%%%%%%%%%%%%%%%

\section{\bf Introduction}

It has recently been proposed \cite{Intriligator:2006dd} that susy-breaking in a
meta-stable vacuum is an interesting paradigm for model-building. The idea
is that in field space, far away from the supersymmetric vacuum, we
could have local minima of the effective potential with non-zero energy.
These vacua can have parametrically long lifetimes because they are
protected by tunneling, so using them as phenomenologically viable
candidates for susy-breaking is a natural possibility. Even simple
models seem to exhibit the existence of such meta-stable vacua, so it
is reasonable to hope that this is a rather generic feature of
supersymmetric theories. Since the original
paper came out, many authors have tried to extend this idea in many
directions \cite{Franco:2006es,Garcia-Etxebarria:2006rw,Ooguri:2006pj,Kitano:2006wm,Braun:2006em,Braun:2006da,Banks:2006ma,Franco:2006ht,Forste:2006zc,Schmaltz:2006qs,Correia:2006vf,Amariti:2006vk,Bena:2006rg,Ahn:2006gn}.

\begin{figure}[b]
\subfigure[\label{T1} Effective potential for $T\gg
T_{c}$]{\includegraphics[width=7cm,height=7cm,keepaspectratio]{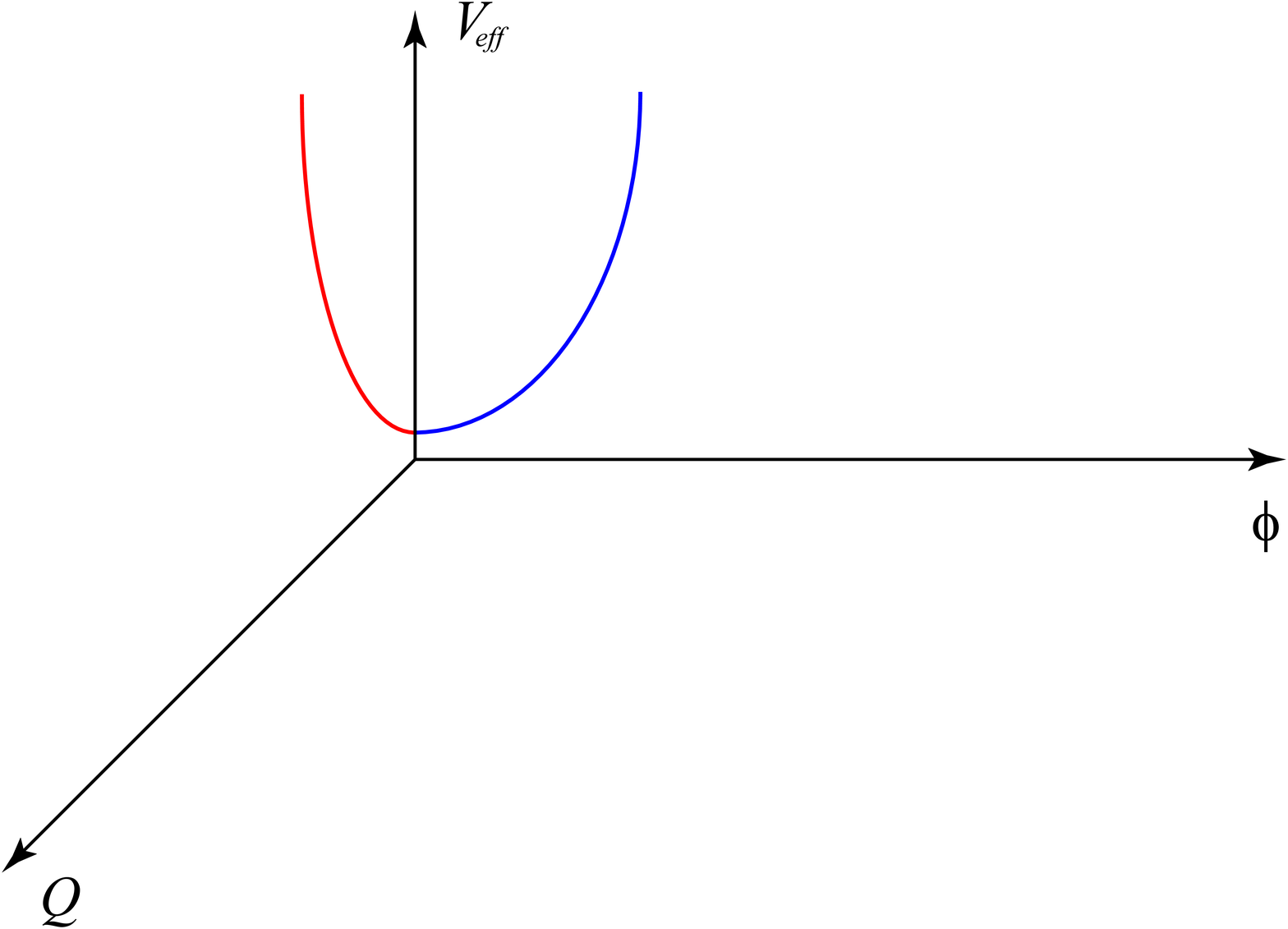}}\subfigure[\label{T2}
Effective potential for $T\sim
T_{c}$]{\includegraphics[width=7cm,height=7cm,keepaspectratio]{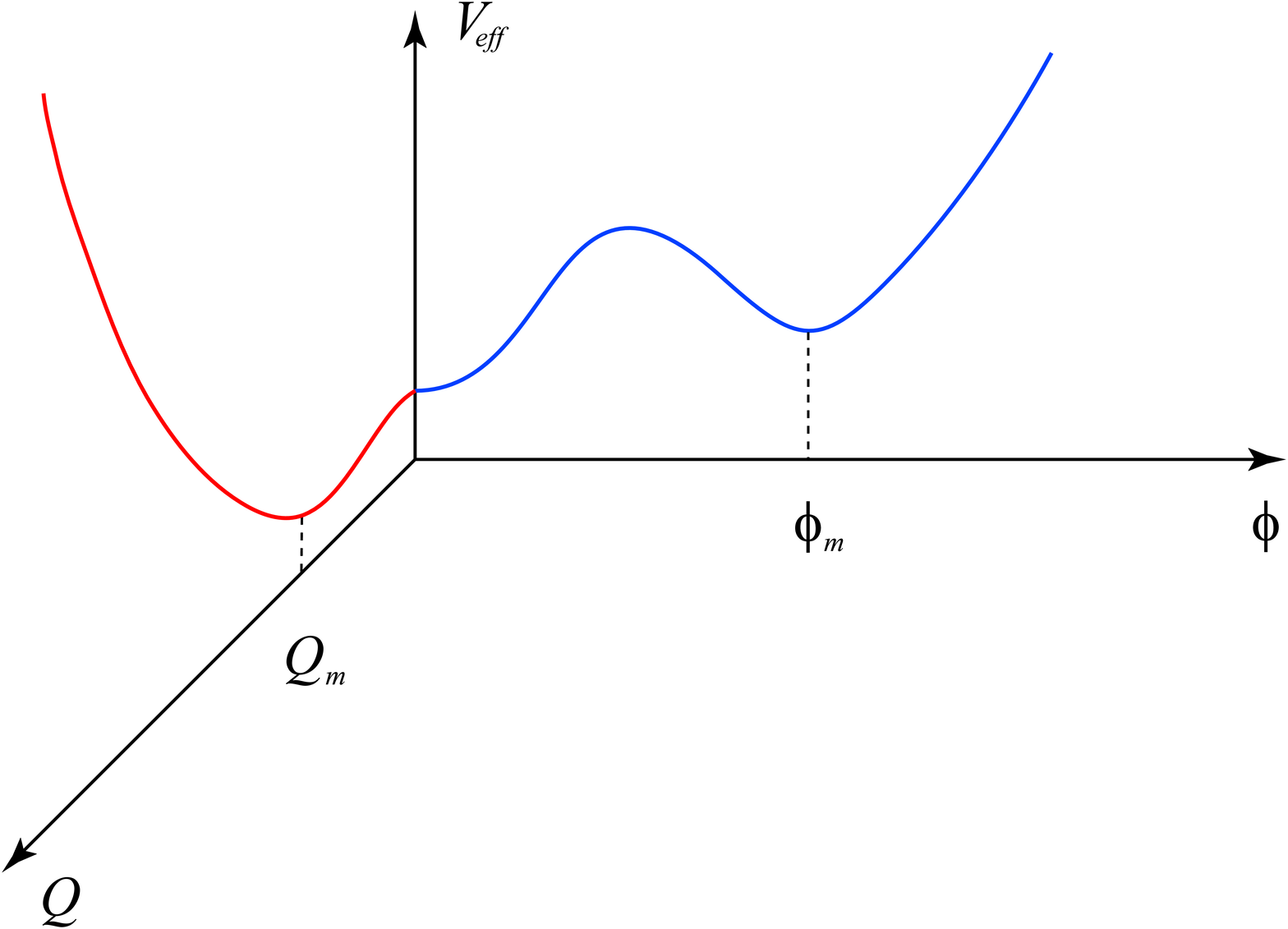}}
\subfigure[\label{T3} Effective potential for
$T=0$]{\includegraphics[width=7cm,height=7cm,keepaspectratio]{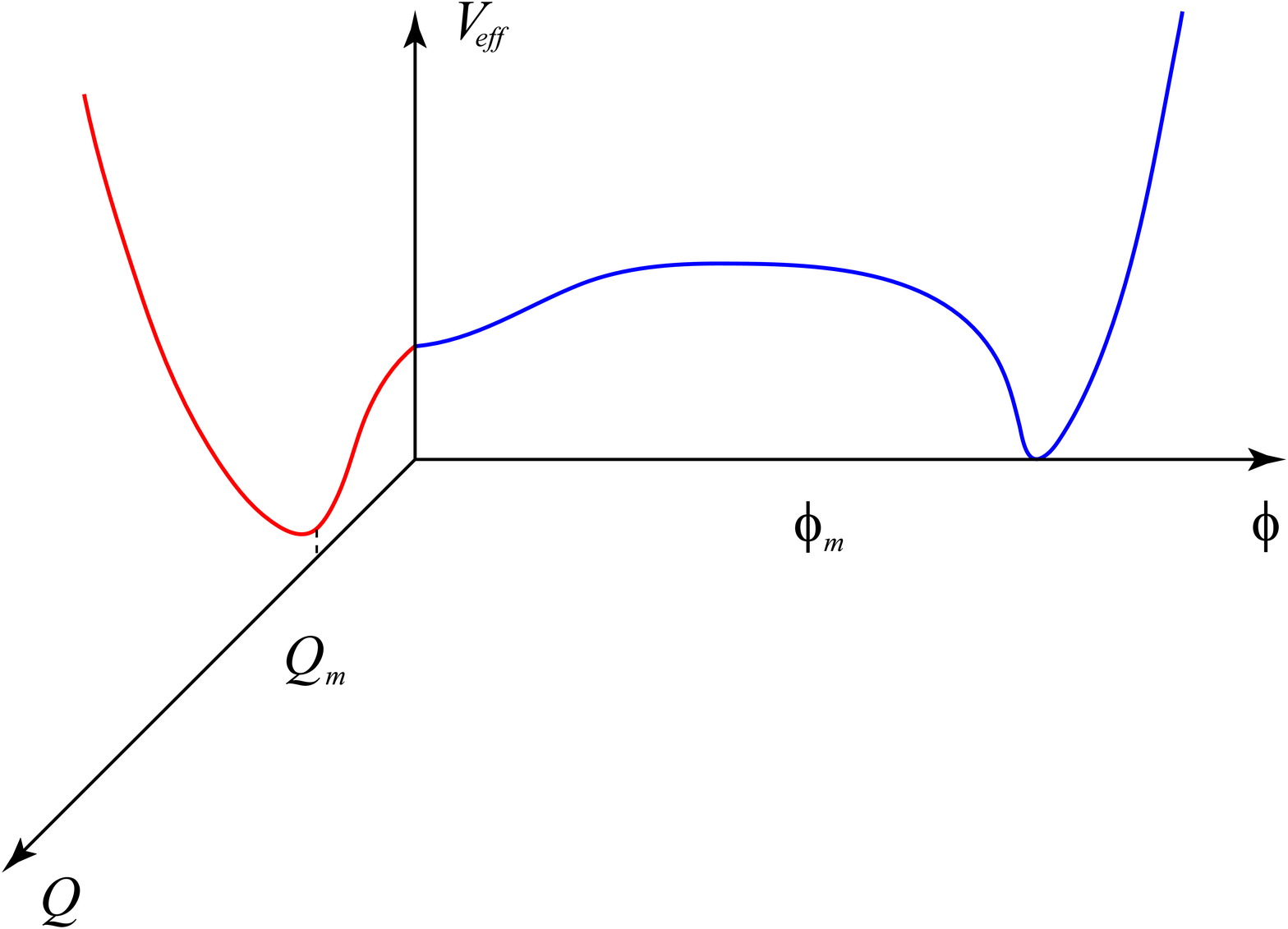}}\caption{Evolution
of the effective potential with temperature}
\end{figure}

In this paper, we will be interested in this question from a cosmological
perspective. Just the fact that there exists a meta-stable vacuum in the
zero-temperature field theory, is not enough to guarantee that we will
end up in it (and {\em not} in the susy-vacuum), as the Universe cools
from high-temperatures. One way to get a better understanding of this
issue is to look at the phase structure of the free energy at different
temperatures. This is precisely the
purpose of this paper.
We work in the context of one of the simplest models
presented in \cite{Intriligator:2006dd}, namely the
S-dual of SuperQCD\footnote{with appropriately chosen number of colors and
flavors.},
and look at the theory at finite temperature. The scalars in the theory
are dual squarks and mesons. At zero-temperature, this theory has a
supersymmetric vacuum  for large values of the meson
field. This is the usual susy-vacuum of SQCD, as seen from the dual
picture.
But the interesting thing about the dual picture is that here, we can
identify new meta-stable susy-breaking vacua of the effective potential
which are invisible from the pure SQCD picture. This was the key insight
of \cite{Intriligator:2006dd}. A schematic picture of the effective
potential at zero
temperature is provided in figure \ref{T3}.

How does this scenario change when we turn on temperature? We can get some
idea about the situation by looking at the equilibrium thermal properties
of this theory at finite temperature
\cite{Kirzhnits:1974as,Weinberg:1974hy,Dolan:1974gu,Coleman:1973jx}. We
could calculate the free
energy
as a function of the quark and the meson fields. At high enough
temperature,
we expect that free energy has a trough at the origin of field space,
because the fields are massless there and the entropy is therefore higher.
We could calculate the mass-matrices for the fields in the quark and
meson directions to get an idea about the free-energy in those
directions. Indeed, when we do this,
we end up finding that as the temperature is lowered,
the second order phase transition in the quark direction happens first.
There is no (second order) phase transition in the meson direction, all
the way down to $T\sim 0$. From these, we conclude that as the
temperature drops, the Universe winds up in a susy-breaking phase.

This argument implies that the meta-stable vacuum is plausible, but
we should add that this in itself is not completely conclusive. The
free energy is a purely equilibrium quantity and the dynamics of the
fields as they interact with a thermal bath could be more
complicated. To fully understand the situation, we need to do a
calculation that incorporates finite temperature effects {\em and}
dynamics. We need to use the evolution of the field at finite
temperature: we could use the imaginary-time formulation and
calculate the friction-type term for the field-equation in the
thermal bath. We do not report on this calculation here, but the
punch-line of the preliminary calculation is that at least for some
initial values of the fields, we do evolve and end up in the
susy-breaking vacuum. The free energy equations give us an idea about
this phase structure, as is clear from the pictures below. In each picture
two separate situations
are plotted together: one in the scalar mesons direction with zero
squark vev and another in the squark direction with zero
scalar meson vev. In \ref{T1}, at high temperature the free energy
drives the scalar expectation values to zero. In \ref{T2}, the plot is at
lower
temperatures close to a threshold where the free energy in the
squark direction develops a minimum away from zero, while in the
meson direction it develops a potential barrier. Finally in \ref{T3}, at
zero temperature, the meta-stable vacuum in the squark direction
remains cosmologically stable \cite{Intriligator:2006dd} due to the
large potential barrier in the scalar meson direction, where the
susy vacuum is located.

One important assumption is that the Seiberg transition is a
post-inflation phenomenon. The reheating temperature is higher than
the Seiberg transition; from the phase where the weakly coupled
degrees of freedom are that of SQCD (electric phase) to the phase
where the weakly coupled degrees of freedom are that of its Seiberg
dual (magnetic phase).  Besides, the unsolved problem of initial
value of the scalar mesons after the Seiberg transition
%allied to the negleted
%dynamics of the fields
allow other possibilities that are not considered in the present
work.

 If initially, the scalar mesons  are located far away in field
space, they will execute large amplitude oscillations. Then,  if
both the Hubble friction and the interactions of the meson fields
with the heat bath are small enough, these oscillations might bring
and keep the meson fields in the vicinity of the susy vacuum.

With these assumptions the scalar fields are not in equilibrium with
the rest of the universe and therefore we should not rely on an
equilibrium calculation of the free energy  to describe this system.
It is then plausible, that as the universe cools, the meson fields
find themselves trapped in the susy state. However in an adiabatic
evolution of the universe it is appropriate to use thermodynamics of
equilibrium  to describe the system. This is what the body of this
paper will make explicit.

In the next section, we introduce the model under consideration
as it was described in \cite{Intriligator:2006dd}. In section 3, we
begin the calculation of the equilibrium free energy of this theory along
the quark and
meson directions
by calculating the mass matrices for the appropriate fields and
expressing the free energy in the high-temperature
limit. The interpretation of this
object as the
temperature is lowered is the subject of section 5.
We draw some conclusions
based on our results and end with some possibilities for future work in
section 6.

\section{\bf Vacuum Structure of the Magnetic Dual of SQCD}

The meta-stable susy breaking vacuum is in a region of field space which
is easier seen in the IR (magnetic) dual of SQCD. We report some of the
results regarding this theory (following the original paper) in this
section, but this is by no means an exhaustive review.

The superpotential for the theory (including the non-perturbative
piece arising from gaugino condensation) is,
\eqn{superpotential}{W=h \ {\rm Tr} \ q\Phi \tilde q-h\mu^2 \ {\rm
Tr} \Phi+ AN( {\rm det} \Phi)^{1/N}} where \eqn{defineA}{A\equiv
h^{\nu}{\Lambda_m}^{-\nu+3}, \ \ \nu= N_f/N.} The matter content can
be described by (the columns denote the $SU(N)$ gauge and global
symmetry groups):

        \begin{center}
                \begin{tabular}{ccccccc}
                        \ &$SU(N)$& $SU(N_f)$&$SU(N_f)$&$U(1)_B$&$U(1)'$&$U(1)_R$\\[1ex]
                        $\Phi$&$1$&$\Box$&$\overline{\Box}$&$0$&$-2$&$2$\\[1ex]
                        $q$&$\Box$&$\overline{\Box}$&$1$&$1$&$1$&$0$\\[1ex]
                         $\tilde q$&$\overline{\Box}$&$1$&$\Box$&$-1$&$1$&$0$ \\[1ex]
                \end{tabular}
        \label{tab:fermionMasses}
\end{center}

The second piece in the superpotential breaks the $SU(N_f)\times SU(N_f)$
down to a diagonal $SU(N_f)$ subgroup.
In our notation, the $q$ stand for the (dual) quarks, the $\Phi$ are gauge
singlet mesons and $\Lambda_m$ is the dynamically generated scale (the
scale of the
Landau pole) for the IR theory. The number of colors of the magnetic
theory $N$, and the number of flavors $N_f$, satisfy $N_f>3N$
so that the
theory is IR free. Notice that all our notation is defined in terms of
the dual theory and {\it not} directly in terms of the microscopic theory
(which is SQCD).

For small meson fields, the non-perturbative term can be dropped, and we
can calculate the moduli space of susy-breaking tree-level vacua. These
are at
\eqn{modulispaceofvacua}{\Phi=\begin{pmatrix}0&0\cr 0&\varphi
\end{pmatrix}, \ \ q=\begin{pmatrix}Q \cr 0 \end{pmatrix}, \ \ \tilde
q=\begin{pmatrix} \tilde Q&0 \end{pmatrix},} with $\tilde Q Q=\mu^2
\unit_{N}$ where $\tilde Q$  and $Q$ are $N\times N$ matrices and
$\varphi$ is a $(N_f-N)\times (N_f-N) $ matrix. In this vacum the
scalar potential has the value
\[V_{min} = (N_f-N)|h^2\mu^4|\]
 The point of maximum global symmetry in this moduli space, up to gauge
transformation and flavor rotation, is
\eqn{maximalsymmetry}{\varphi=0, \ \ Q=\tilde Q =\mu \unit_{N},}
% and at zero temperature, it is these enhanced symmetry vacua that are
%(meta-)stabilized by loop corrections.
 The interesting point is that
along with these meta-stable vacua, far away in the moduli-space at
large meson values, we have supersymmetric vacua that arise from the
extremization of the superpotential. These are of course the
familiar SUSY vacua of SQCD, seen from the dual picture.

%\begin{figure}
%\centering{\includegraphics{metavacuum3}}
%\caption{Schematic vacuum structure of zero-temperature theory}
%\end{figure}

Putting all these together, the vacuum structure of the IR dual of
SQCD looks schematically like Fig. \ref{T3}. The susy-breaking
vacuum is in a local trough (in all directions, even though we
indicate only the quark direction), so it is meta-stable and it is
cosmologically stable due to a large potential barrier.

\section{Free Energy}

Our aim now, is to look at the above theory, and calculate the finite
temperature effective potential (free energy) up to one loop. We want to
see how the picture above changes when we turn on temperature. This will
give us a clue about the possible phase transitions that could happen, as
the universe cooled. The standard procedure for calculating the
finite-temperature effective potential is to shift the
relevant background fields (in our case the scalar quarks and mesons) and
use the resulting quadratic pieces in the action to do the computation. So
essentially, we
need to know what the masses\footnote{To be more precise, we should
say that we are after the coefficients of the pieces in the action that
are quadratic in fields, we will call them masses even though we are not
necessarily looking at the extremum of the potential.} of
the various fields are, as a result of the shifts in the backgrounds. It
is the calculation of these mass matrices, that we undertake next. For the
sake of simplicity, we will take the parameters $h$ and $\mu$ as well
as the shifts in the background fields to be real numbers.

\subsection{\bf Mass Matrices in the Meson Directions}

We first calculate the masses of the various fields when a background
field is turned on in the meson direction, with the scalar quarks set to zero. In the
next subsection we will turn on the squarks and turn off the mesons.
Once we know the mass matrices in both these directions, we can draw some
conclusions about the free energy and the phase structure as a function of
temperature.

Decomposing the meson chiral superfield into a background, trace and a
traceless part,
\eqn{phisplit}{\Phi = \Big(\varphi+\frac{\Phi_0}
{\sqrt {N_f}}\Big)\unit +\widehat \Phi,}
we can expand the nonperturbative term in the superpotential:
\begin{align*}
 AN(\det\Phi)^{1/N} = AN\varphi^\nu\exp\left(\tfrac{1}{N}{\rm Tr}
\ln(\unit +
\tfrac{\Phi_0}{\sqrt {N_f}\varphi}\unit+\tfrac{\widehat
\Phi}{\varphi})\right)\sim \\
 AN \varphi^\nu\exp\left(\tfrac{1}{N}(\tfrac{\Phi_0\sqrt
{N_f}}{\varphi}-\tfrac{\Phi_0^2}{2\varphi^2}-
\tfrac{{\rm Tr}(\widehat \Phi^2)}{2\varphi^2}+\tfrac{\Phi_0^3}{3\sqrt
{N_f}\varphi^3})\right)\sim\\
 A \varphi^\nu\left(N+  \frac{\Phi_0\sqrt {N_f}}{\varphi}+
\frac{(\nu-1)\Phi_0^2}{2\varphi^2}-\frac{{\rm Tr}(\widehat
\Phi^2)}{2\varphi^2}+
\frac{\Phi_0^3(\nu-1)(\nu-2)}{6\sqrt {N_f}\varphi^3}-
\frac{(\nu-2)\Phi_0 {\rm Tr} (\widehat \Phi^2)}{2\sqrt
{N_f}\varphi^3}\right)
\end{align*}
>From this, we can read off the terms that contribute to the fermion
masses in the Lagrangian:
\eqn{ferm-mass-meson}{\mathcal{L}\supset h\varphi \ {\rm
Tr}(\psi_q\psi_{\tilde
q})+
\frac{A\varphi^{\nu-2}(\nu -1)}{2}\psi_{\phi_0}\psi_{\phi_0}-
\frac{A\varphi^{\nu-2}}{2}\ {\rm Tr}(\psi_{\hat
\phi}\psi_{\hat\phi})+h.c.}
showing that there are $N\times N_f$ Dirac fermions
with mass $h\varphi$; 1 Majorana fermion with mass
$A\varphi^{\nu-2}(\nu -1)$ and $(N_f^2-1)$ Majorana fermions with
mass $A\varphi^{\nu-2}$.

\begin{table}
    \centering
        \begin{tabular}{|c|c|c|}\hline
            \# of weyl fermions& fermion fields & mass
\\ [1ex] \hline
            $2N\times N_f$&$\psi_q ,\; \psi_{\tilde q}$&$h\varphi$\\[.5ex] \hline
                $1$&$\psi_{\phi_0} $&$A\varphi^{\nu-2}(\nu -1)$\\[.5ex] \hline
                $(N_f^2-1)$&$\psi_{\hat \phi}$&$A\varphi^{\nu-2}$\\[.5ex] \hline
        \end{tabular}
    \caption{fermions masses, meson direction}
    \label{tab:fermionMasses}
\end{table}

To calculate the scalar masses from the scalar potential, we first
compute the F-terms.
\begin{eqnarray} F_{\phi_0}&=&\frac{h}{\sqrt {N_f}} \ {\rm Tr}(q\tilde q)
-h\mu^2
\sqrt {N_f} +A\varphi^\nu\left(\frac{\sqrt {N_f}}{\varphi} + \frac{\nu-1}{\varphi^2}\phi_0 + \frac{(\nu-1)(\nu-2)}{2\varphi^3\sqrt {N_f}}\phi_0^2\right)\\
F_{\hat \phi}&=& h\tilde q q + A\varphi^\nu\left(-\frac{\widehat \phi^t}{\varphi^2}-\frac{(\nu-2)}{\sqrt {N_f}\varphi^3}\phi_0 \widehat \phi^t\right)\\
F_q&=&h(\phi\tilde q)^t\supset h(\varphi \tilde q)^t\\
F_{\tilde q}&=&h(\phi q)^t\supset h(\varphi q)^t
\end{eqnarray}
Consequently, the scalar potential will have the following quadratic terms:
\begin{gather}V_{scalar}\supset (A\varphi^{\nu-1}-h\mu^2)\left[
{\rm Tr}(q\tilde
q)h+\frac{(\nu-1)(\nu-2)A\varphi^{\nu-3}}{2}\phi_0^2\right]+c.c.+\\
|A|^2\varphi^{2\nu-4}((\nu-1)^2|\phi_0|^2+|\widehat \phi|^2)+
h^2\varphi^2(|\tilde q|^2+|q|^2)
\end{gather}
where the scalar masses are extracted and shown in table \ref{tab:scalarSquaredMassesmesondirection}.

\begin{table}
    \centering
        \begin{tabular}{|c|c|c|} \hline
            \# of real scalars& scalar fields &
$mass^2$\\[1ex]
\hline
            $2(N_f^2-1)$&$\widehat
\phi$&$|A|^2\varphi^{2\nu-4}$\\[.5ex] \hline
            $1$&$Im \
\phi_0$&$\begin{tabular}
{c}$|A|^2\varphi^{2\nu-4}(\nu-1)^2-$\\
$-Re\left[(A\varphi^{\nu-1}-h\mu^2)^*(\nu-1)(\nu-2)A\varphi^{\nu-3}\right]$\end{tabular}$\\[.5ex] \hline
            $1$&$Re \ \phi_0$&$\begin{tabular}
{c}$|A|^2\varphi^{2\nu-4}(\nu-1)^2+$
\\
$+Re\left[(A\varphi^{\nu-1}-h\mu^2)^*(\nu-1)(\nu-2)A\varphi^{\nu-3}\right]$
\end{tabular}$\\[.5ex] \hline
            $N\times N_f$&$Re(q+\tilde q^t)/\sqrt
2$&$h^2\varphi^2+Re(A\varphi^{\nu-1}-h\mu^2)h$\\[.5ex]\hline
            $N\times N_f$&$Im(q+\tilde q^t)/\sqrt
2$&$h^2\varphi^2$\\[.5ex]\hline
            $N\times N_f$&$Re(q-\tilde q^t)/\sqrt
2$&$h^2\varphi^2-Re(A\varphi^{\nu-1}-h\mu^2)h$\\[.5ex]\hline
            $N\times N_f$&$Im(q+\tilde q^t)/\sqrt
2$&$h^2\varphi^2$\\[.5ex]\hline
        \end{tabular}
    \caption{real scalar squared masses, meson direction}
    \label{tab:scalarSquaredMassesmesondirection}
\end{table}

\subsection{\bf Mass Matrices in the Quark Directions}

The mass matrices in the quark directions are more complicated than in the
meson directions because there are contributions from the D-terms. We
start by classifying the various sectors according to their
transformation properties under the global symmetries. We take the
shifts in the form:
 $ \langle\tilde q_1\rangle=\langle q_1\rangle =Q\unit $, $Q$ real.
The column vectors are $N \times N_e$ where $N_e=N_f-N$. The subscript
stands for electric. $N_e$ is the same as $N_c$, the number of colors
in the original (microscopic) theory, namely SQCD, but we prefer to think
of it purely in terms of the dual theory. On a similar note, $N$ is
actually $N_m$, with the subscript standing for magnetic.

There are four sectors that do not mix with each other in the Lagrangian.
We will use this fact to our advantage in calculating the mass matrices.
These sectors are:

1. $N_e \times N_e$ : $\phi_{22}$,

2. $N_e \times N_m$ : $\tilde q_2, \phi_{21}$,

3. $N_m \times N_e$ : $q_2, \phi_{12}$,

4. $N_m \times N_m$ : $q_1, \tilde q_1, \phi_{11}, V$.

The $V$ in the last line is the vector superfield, it gets massive
through a Higgs mechanism.

Now we look at the masses of the various fields sector by sector. To start
off, in the $ N_e \times N_e$ sector, there are $N_e^2 \times ({\rm
complex \ scalars} + {\rm Weyl \ Fermions})$, and all of them remain
entirely massless (at tree level). But it is in this sector that
supersymmetry is broken by a positive contribution in the scalar
potential. Decomposing $\Phi_{22}$ into trace and traceless parts, we get
\[ \Phi_{22}=\frac{\Phi_{22}^0}{\sqrt{N_e}}\unit+\widehat
\Phi_{22};\;\;F_{\phi_{22}^0}\supset-h\mu\sqrt{N_e}.\]
Hence
\[V_{scalar}\geq N_eh^2\mu^4.\]

It is easiest to deal with the two mixed electric/magnetic sectors (2 and
3) together.
Together, there are $2 N_e N_m \times (2 \ {\rm complex \ scalars} + 1\
{\rm Dirac \ fermion})$. The fermionic
masses arise from terms like
\eqn{2and3}{hQ \ {\rm Tr}(\Psi^\phi_{12}\Psi^q_2)+hQ \ {\rm
Tr}(\Psi^{\tilde
q}_2\Psi^\phi_{21})+h.c.,}
%The fermion masses arise from the matrix
%\eqn{mf23}{M_f=....}
and these give rise to a Dirac mass of $hQ$. To calculate the bosonic
masses, we need the scalar potential in the quark direction which can be
calculated easily enough from the superpotential and the F-terms. The
F-terms $F_{q_1}, F_{\tilde q_1}$ do not
give rise to scalar masses in sectors 2 and 3 because the scalars from
these sectors have zero vevs.  The relevant non-vanishing ones are
\eqn{Fterm1}{F_{\phi_{12}}=h(\tilde q_2 q_1)^t\supset hQ(\tilde q_2)^t}
\eqn{Fterm3}{F_{\phi_{21}}=h(\tilde q_1 q_2)^t\supset hQ(q_2)^t}
\eqn{Fterm2}{F_{\phi_{22}}=h(\tilde q_2 q_2)^t-h\mu^2\times \unit}
\eqn{Fq2}{F_{\tilde q_2}=h(q_1\phi_{12} +q_2\phi_{22})\supset hQ\phi_{12}}
\eqn{Fq2tilde}{F_{q_2}=h(\phi_{21}\tilde q_1+ \phi_{22}\tilde q_2)\supset
hQ\phi_{21}}
Therefore, the scalar potential contains the quadratic terms:
\eqn{scalarepot-quark}{V_{scalar}\supset h^2 \left(Q^2|\tilde q_2|^2+
Q^2|q_2|^2 - \mu^2 \ {\rm Tr}(\tilde q_2
q_2+h.c.)+Q^2|\phi_{12}|^2+Q^2|\phi_{21}|^2\right)}
where in our notation modulus squared of matrices means trace over the
product of the matrix and its adjoint. As we see,
$2N_m\times N_e$ complex scalars ($\phi_{12}$ and $\phi_{21}$) get
squared mass $h^2Q^2$, $2N_m\times N_e$
real scalars ($Re(q_2\pm\tilde q_2^t)/\sqrt 2$) split
their masses into $h^2(Q^2\pm\mu^2)$ and another
$2N_m\times N_e$ real scalars
 ($Im(q_2\pm\tilde q_2^t)/\sqrt 2$) get
mass $h^2Q^2$. See table (\ref{tab:scalarSquaredMasses23}).

\begin{table}
    \centering
        \begin{tabular}{|c|c|c|} \hline
            \# of real scalars& scalar fields &
$mass^2$\\[1ex] \hline
            $4N_m\times N_e$&$\phi_{12}$ and
$\phi_{21}$&$h^2Q^2$\\[.5ex] \hline
            $N_m\times N_e$&$Re(q_2+\tilde q_2^t)/\sqrt
2$&$h^2(Q^2+\mu^2)$\\[.5ex] \hline
            $N_m\times N_e$&$Re(q_2-\tilde q_2^t)/\sqrt
2$&$h^2(Q^2-\mu^2)$\\[.5ex] \hline
            $N_m\times N_e$&$Im(q_2+\tilde q_2^t)/\sqrt
2$&$h^2Q^2$\\[.5ex] \hline
            $N_m\times N_e$&$Im(q_2-\tilde q_2^t)/\sqrt
2$&$h^2Q^2$\\[.5ex] \hline
        \end{tabular}
    \caption{real scalar squared masses, sectors 2 and 3}
    \label{tab:scalarSquaredMasses23}
\end{table}

Now we turn to sector 4. First, we separate out the
background:
\eqn{qs}{q_1 = Q\unit + \widehat q_1\;;\;\;\tilde q_1 = Q\unit  +\widehat{\tilde q}_1}
%and
%\eqn{phi11}{\phi_{11}= \frac{\phi_{11}^0}{\sqrt{N_m}}\unit + \widehat \phi_{11}}
Some of the fermion masses arise from the terms
\eqn{Yuk_q}{g\sqrt{2}Q\ {\rm Tr}(\lambda\Psi^q_1) - g\sqrt{2}
Q\ {\rm Tr}(\lambda\Psi^{\tilde q}_1) +h.c.}
where  gauginos $\lambda^a$, and the traceless $(\Psi^q_1 - \Psi^{\tilde q}_1)/\sqrt{2}$ have equal masses $2gQ$.

>From the K\"ahler potential, the vector bosons $A_\mu$ get a mass  $2gQ$.
The  traceless part of $Im \left((\widehat {q_1} -
\widehat{\tilde q}_1)/\sqrt{2}\right)$ is gauged away and the
traceless part of the scalars $Re \left((\widehat{q_1}
-\widehat{\tilde q_1})/\sqrt{2}\right)$ get their masses from the
F-terms (to be calculated) and from the D-terms. The contribution
to this squared mass from
the D-terms can be read off from
\begin{eqnarray}
\frac{g^2}{2}\sum_a({\rm Tr} (q^{\dagger}_1t^aq_1-\tilde q_1t^a\tilde q^\dagger_1))^2\supset \frac{g^2}{2}\sum_a( Q{\rm Tr} (t^a(\widehat q_1-\widehat{\tilde q}_1+ \widehat q^\dagger_1- \widehat{\tilde q}^\dagger_1))^2= \notag \\
\frac{g^2}{2}\sum_a( Q{\rm Tr} (2Re[t^a(\widehat q_1-\widehat{\tilde q}_1 )]))^2= 4g^2Q^2\sum_a\left( {\rm Tr} (Re[t^a(\widehat q_1-\widehat{\tilde q}_1 )/\sqrt 2])\right)^2
\label{Dterms}
\end{eqnarray}
to be $4g^2Q^2$. The   scalar $Im\left({\rm Tr}(q_1-\tilde
q_1)\right)/\sqrt{2}$ and the fermion
${\rm Tr}(\Psi^q_1 - \Psi^{\tilde q}_1)/\sqrt{2}$ remain massless at
tree level. But $Re({\rm Tr}(q_1-\tilde q_1))/\sqrt{2}$ receives a contribution
from the F-terms.

The shifts in $q_1$ and $\tilde q_1$ give rise to terms of
the form
\[
hQ\Psi^{\phi}_{11}(\Psi^q_1+\Psi^{\tilde q}_1)+h.c.\]  From these, the
$2N_m^2$ Weyl
fermions $(\Psi^{\phi}_{11})$ and $(\Psi^q_1+\Psi^{\tilde q}_1)/\sqrt{2}$
acquire a mass of $hQ\sqrt{2}$ each.
\begin{table}
    \centering
        \begin{tabular}{|c|c|c|} \hline
            \# fields& fields & mass\\[1ex]\hline
            $N_m^2-1$&$A^{\mu,a}$&$2gQ$\\[.5ex]\hline
            $N_m^2-1$&$\lambda^a$&$2gQ$\\[.5ex]\hline
            $N_m^2-1$& traceless $(\Psi^q_1 - \Psi^{\tilde q}_1)/\sqrt{2}$&$2gQ$\\[.5ex]\hline
            $2N_m^2$&$\Psi^{\phi}_{11}$&$hQ\sqrt 2$\\[.5ex]\hline
            $2N_m^2$&$(\Psi^q_1+\Psi^{\tilde q}_1)/\sqrt{2}$&$hQ\sqrt 2$\\[.5ex]\hline
        \end{tabular}
    \caption{vector boson and fermion masses, sector 4}
    \label{tab:vectorfermionmasses}
\end{table}
Their respective scalar superpartners
acquire mass through the scalar potential. Writing the relevant
terms in the F-terms, using (\ref{qs}), we have
\[F_{\phi_{11}}=h\left((Q^2-\mu^2)\unit +Q\sqrt2\frac{\widehat{q_1}+\widehat{\tilde q_1} }{\sqrt2} +\widehat{\tilde q_1}\widehat{q_1}\right)\]
and $F_{q_1}\supset hQ\phi_{11}\;\; ,\;\;F_{\tilde q_1}\supset
hQ\phi_{11}$. Hence, the quadratic terms in the scalar potential are,
\eqn{scalarpotential}{V_{scalar} \supset
h^2\left(2Q^2\left|\phi_{11}\right|^2+ 2Q^2\left|\frac{\widehat q_1
+\widehat {\tilde q}_1}{\sqrt {2}}\right|^2+
(Q^2-\mu^2)\left(\ {\rm Tr}(\widehat{\tilde q_1}\widehat
q_1)+c.c\right)\right)}

This shows that among the real scalars, $2N_m^2$ get
squared masses $2h^2Q^2$, $N_m^2$ get squared masses  $h^2(3Q^2-\mu^2)$,
and $N_m^2$ get squared masses $2h^2Q^2$. The
corresponding fields are $\phi_{11}$, $Re(\widehat{\tilde q_1}+
\widehat{q_1})/\sqrt 2)$ and $Im(q_1+\tilde q_1)/\sqrt 2$ respectively.
The
terms  $Re(\widehat{\tilde q_1}-\widehat{q_1})/\sqrt2$ get mass from
above and from the D-terms, splitting the field matrix into a trace part
with mass  $h^2(\mu^2-Q^2)$ and $N_m^2-1$ traceless components with mass
$h^2(\mu^2-Q^2)+4g^2Q^2$.
\begin{table}
    \centering
        \begin{tabular}{|c|c|c|}\hline
            \# of real bosons&
fields &
$mass^2$\\[1ex]
\hline
            $N_m^2$&$Re \ (\widehat{q_1}+\widehat{\tilde
q}_1)/\sqrt 2$&$h^2(3Q^2-\mu^2)$\\[.5ex]\hline
            $2N_m^2$&$\phi_{11}$&$2h^2Q^2$\\[.5ex]\hline
            $N_m^2$&$Im(\widehat{q_1}+\widehat{\tilde
q}_1)/\sqrt 2$&$2h^2Q^2$\\[.5ex] \hline
            $N_m^2-1$&traceless
$Re(\widehat{q_1}-\widehat{\tilde q}_1)/\sqrt
2$&$h^2(\mu^2-Q^2)+4g^2Q^2$\\[.5ex] \hline
            $1$&${\rm
Tr}\left(Re(\widehat{q_1}-\widehat{\tilde
q}_1)/\sqrt2\right)$&$h^2(\mu^2-Q^2)$\\[.5ex] \hline
            $1$&${\rm
Tr}\left(Im(\widehat{q_1}-\widehat{\tilde
q}_1)/\sqrt2\right)$&$0$\\[.5ex] \hline
%           $3(N_m^2-1)$&$A^{\mu}$&$4g^2Q^2$\\[0.5ex] \hline
        \end{tabular}
    \caption{real squared masses for bosons, sector 4}
    \label{tab:scalarSquaredMasses}
\end{table}

The vector boson, gaugino and fermion masses are presented in table
(\ref{tab:vectorfermionmasses}), and the scalar masses are in
table (\ref{tab:scalarSquaredMasses}).

\subsection{\bf Effective potential}

The effective potential at finite temperature is the free energy of a
system in a thermal bath. In thermal equilibrium, there is no time
dependence and the different phases correspond to
local minima of the free energy.

The finite temperature effective potential up to one loop is given
by \cite{Dolan:1974gu}
\[ V(\phi_{cl})=  V_{tree}(\phi_{cl})+ V_1^0(\phi_{cl}) +
V_1^T(\phi_{cl})\]
where $V_{tree}(\phi_{cl})$ is the classical piece. The
one-loop correction, $V_1^T(\phi_{cl})$,
for a generic theory is:
\begin{gather}
 V_1^T(\phi_{cl})\approx
-\frac{\pi^2T^4}{90}\left(N_B+\frac{7}{8}N_F\right)+\frac{T^2}{24}\left[\sum_i{(M_S^2)_i}+
3\sum_i{(M_V^2)_i} + \sum_i{(M_F^2)_i}\right]\notag \\
 -\frac{T}{12\pi}\left[\sum_i{(M_S^3)_i}+3\sum_i {(M_V^3)}_i\right]+...
 \label{V1T}\end{gather}
and $V_1^0(\phi_{cl})$ is the zero-temperature piece,
$(M_S)_i,\;(M_V)_i$ and $(M_F)_i$ rae mass-matrix eigenvalues for
the real scalars, vectors and Weyl fermions respectively, and $N_B =
N_F$ is the number of bosonic/fermionic degrees of freedom, paired
by supersymmetry. What we have done here is to follow the standard
practice and split off the one-loop, finite-temperature effective
potential into a part that is independent of temperature (and
therefore is the same as the zero-temperature effective potential)
and then do a high-temperature ($T\gg$ masses) expansion on the
remaining (temperature-dependant) piece. %For convenience, we will
%call this temperature dependant piece $\bar V_1^T(\phi_{cl})$:
%\eqn{vbar}{\bar V_1^T(\phi_{cl})\equiv
%V_1^T(\phi_{cl})-V_1^0(\phi_{cl}),} so that
%\[ V(\phi_{cl})=  V_{tree}(\phi_{cl})+ V_1^0(\phi_{cl}) +
%\bar V_1^T(\phi_{cl}).\]
 The zero temperature piece is calculated
using the usual supersymmetric generalization of the
Coleman-Weinberg formula \cite{Coleman:1973jx}:
 \eqn{CWgen}{V_1^0 =
\frac{1}{ 64\pi^2}{\rm
STr}\,\mathcal{M}^4\log\frac{\mathcal{M}^2}{\Lambda ^2},} where
$\mathcal{M}$ stands for the full mass-matrix, with the supertrace
providing the negative sign for the fermionic terms.

Armed with the above expressions and the mass-matrices from the last
section, its easy to calculate the finite-temperature effective
potential. In the following, we only keep the terms that are quartic and
quadratic in temperature. In the case of
background fields in the meson direction,
we find:
\eqn{Vmtree}{V_{tree}(\varphi) = N_f|(A\varphi^{\nu-1}-h\mu^2)|^2,}
\begin{gather} V_1^0(\varphi) =
\frac{1}{64\pi^2}\Big(-2|A|^4\varphi^{4\nu-8}(\nu-1)^4
\log\frac{|A|^2\varphi^{2\nu-4}(\nu-1)^2}{\Lambda^2}+\notag
\\ +[|A|^2\varphi^{2\nu-4}(\nu-1)^2-
Re\left[(A\varphi^{\nu-1}-h\mu^2)^*(\nu-1)(\nu-2)A
\varphi^{\nu-3}\right]]^2\times
\notag \\
\times\log\frac{|A|^2\varphi^{2\nu-4}(\nu-1)^2-Re\left[(A\varphi^{\nu-1}-
h\mu^2)^*(\nu-1)(\nu-2)A\varphi^{\nu-3}\right]}{\Lambda^2}+\notag
\\
+[|A|^2\varphi^{2\nu-4}(\nu-1)^2+
Re\left[(A\varphi^{\nu-1}-h\mu^2)^*(\nu-1)(\nu-2)A\varphi^{\nu-3}
\right]]^2\times
\notag \\
\times\log\frac{|A|^2\varphi^{2\nu-4}(\nu-1)^2+Re\left[(A\varphi^{\nu-1}-
h\mu^2)^*(\nu-1)(\nu-2)A\varphi^{\nu-3}\right]}{\Lambda^2}+\notag\\
+ N_mN_fh^2(h\varphi^2+Re(A\varphi^{\nu-1}-h\mu^2))^2
\log\frac{h^2\varphi^2+Re(A\varphi^{\nu-1}-h\mu^2)h}{\Lambda^2}-\notag \\
            -2N_m
N_fh^4\varphi^4\log\frac{h^2\varphi^2}{\Lambda^2}+
N_mN_fh^2(h\varphi^2-Re(A\varphi^{\nu-1}-h\mu^2))^2
\log\frac{h^2\varphi^2-
Re(A\varphi^{\nu-1}-h\mu^2)h}{\Lambda^2}\Big),\label{Vm0}\end{gather}
\begin{gather} \bar V_1^T(\varphi) =
-\frac{\pi^2T^4}{24}\left((N_f+N_m)^2-1\right)+
\notag \\
+\frac{T^2}{24}[3(N_f^2-1)|A|^2\varphi^{2\nu-4}+
3|A|^2\varphi^{2\nu-4}(\nu-1)^2 +8N_mN_fh^2\varphi^2].
\label{VmT}
%-\frac{T}{12\pi}\left[\sum_i{(M_S^3)_i}+3\sum_i {(M_V^3)}_i\right].
\end{gather}
Similarly, for the quark direction:
\eqn{Vqtree}{V_{tree}(Q) = N_eh^2\mu^4+N_mh^2(Q^2-\mu^2)^2,}
\begin{gather} V_1^0(Q) =\frac{1}{64\pi^2}(-2N_m
N_eh^4Q^4\log\frac{h^2Q^2}{\Lambda^2}+N_m N_eh^4
(Q^2+\mu^2)^2\log\frac{h^2(Q^2+\mu^2)}{\Lambda^2}+ \notag \\
+N_m N_eh^4(Q^2-\mu^2)^2\log\frac{h^2(Q^2-\mu^2)}{\Lambda^2}+
N_m^2h^4(3Q^2-\mu^2)^2\log\frac{h^2(3Q^2-\mu^2)}{\Lambda^2}-\notag \\
-4N_m^2h^4Q^4\log\frac{2h^2Q^2}{\Lambda^2}+(N_m^2-1)(h^2(\mu^2-Q^2)+4g^2Q^2)^2\log\frac{h^2(\mu^2-Q^2)+4g^2Q^2}{\Lambda^2}+\notag
\\
+(h^2(\mu^2-Q^2))^2\log\frac{h^2(\mu^2-Q^2)}{\Lambda^2}-
(N_m^2-1)(4g^2Q^2)^2\log\frac{4g^2Q^2}{\Lambda^2}),
\label{Vq0}\end{gather}
\begin{gather}\bar V_1^T(Q) =-
\frac{\pi^2T^4}{24}\left((N_f+N_m)^2-1\right)+
\frac{T^2Q^2}{6}\left[3N_mN_eh^2+2N_m^2h^2+g^2+5(N_m^2-1)g^2\right].
\label{VqT}\end{gather}

With these explicit forms for the finite temperature effective potential,
we will be able to draw some conclusions about the nature of the
phase-transitions in the next section.

Armed with these expressions and the mass-matrices from the previous
seb-sections, we can calculate the explicit forms for the free energy.

\section{Cooling and the Emergence of Different Phases}

We want now to understand what happens to the mesons and squarks
$\varphi, Q$ during the evolution of the universe, as we cool down
from a high temperature.
%\begin{center}%
%\begin{figure}[b]
%\subfigure[\label{T1} Effective potential for $T\gg T_{c}$]{\includegraphics[width=7cm,height=7cm,keepaspectratio]{T1}}\subfigure[\label{T2} Effective potential for $T\sim T_{c}$]{\includegraphics[width=7cm,height=7cm,keepaspectratio]{T2}}
%\subfigure[\label{T3} Effective potential for $T=0$]{\includegraphics[width=7cm,height=7cm,keepaspectratio]{T3}}\caption{Evolution of the effective potential with temperature}
%\end{figure}
%\par\end{center}
In particular, we want to know the phase structure. If the phase
transition in the quark direction happens at a higher temperature
than in the meson direction, then we have at least some reason to
believe that we will eventually end up in the susy-breaking phase.

Phase transitions are characterized by a critical temperature $T_{c}$.
By definition, the critical temperature $T_{c}$ for a second order
phase transition is the temperature at which the second derivative
of $V(\varphi,Q,T)$ at the origin, in one of the field directions,
changes sign from positive to negative. When this happens, the local
minimum (at the origin) in that direction becomes a local maximum and a
new minimum forms at some finite field value. As a consequence, the vacuum
at the origin becomes unstable, a phase transition takes place, and the
fields evolve to the newly formed minimum. Of course, again, we emphasize
that we are doing an equilibrium analysis, but we believe that this is
enough to give a preliminary, heuristic picture of the field history.
%Given this definition, from the expression for $V(\varphi,Q,T)$,
%(\ref{Vqtree}), (\ref{Vq0}), (\ref{VqT}),
We find that $T_{c}^{Q}$ in the quark direction is given by

\[(T_{c}^{Q})^2=\frac{12\mu^{2}}{3N_{e}+2N_{m}+
\frac{g^{2}(1+5(N_{m}^{2}-1))}{h^{2}N_{m}}}.\]

%\[(T_{c}^{Q})^2=\frac{3h^2\mu^{2}[4N_m+8h^2(N_m^2+1)log{\tfrac{h^2\mu^2}{\Lambda}} + 8h^2N_m^2 -8g^2(N_m^2-1)]}{3N_{e}+2N_{m}+
%\frac{g^{2}(1+5(N_{m}^{2}-1))}{h^{2}N_{m}}}.\] 

%And for $N_m^2>>1$, 
%\[(T_{c}^{Q})^2\approx\frac{8h^2\mu^{2}(\nu-1)(1-\tfrac{g^2}{h^2}+\log \tfrac{h^2\mu^2}{\Lambda})}{1+\tfrac{2(\nu-1)}{3}+\tfrac{5(\nu-1)}{3}\tfrac{g^2}{h^2}}.\] 
%where the relation $g^2/h^2$ is kept fixed by the renormalization group.

On the other hand, in
the meson direction, we see that there is a minimum away from the
origin at this temperature, but, the local minimum at the origin is
still there. At some temperature $T_{c}^{\varphi}$ these two minima
will become degenerate
($V(0,T_{c}^{\varphi})=V(\varphi_{m},T_{c}^{\varphi})$), but there
still is a potential barrier between them. Tunneling through the
barrier can start when the temperature hits $T_{c}^{\varphi}$, but
this phase transition is first order as opposed to the second-order
phase transition in the quark direction. The critical temperature
for the first order phase transition in the mesons direction turns
out to be

\[
(T_{c}^{\varphi})^{2}\sim\left(\frac{24N_{f}}{\pi^{2}\left(\left(N_{f}+
N_{m}\right)^{2}-1\right)}\right)^{\frac{1}{2}}h\mu^{2}+O(h).
\]

%\[
%T_{c}^{\varphi}\sim\sqrt{3/N_m}\left(1 +\tfrac{N_mh^2}{16\pi^2}log(\tfrac{\mu h}{|\Lambda|})\right)^{1/2}h\mu \left(\tfrac{\mu}{|\Lambda|}\right)^{\frac{\nu-3}{\nu-1}}.
%\]

We notice that in the meson direction the origin is always a local
minimum for every temperature, even zero temperature. In fact
expanding the 1-loop effective potential at zero temperature
$V(\varphi,Q,T=0)$ %(\ref{Vqtree}
 around the origin we find \[
V(\varphi,Q,T=0)\sim h^{2}\varphi^{2},\:\varphi\sim0.\]
 Finite temperature effects do not change the fact the the origin
is still a local minimum in the meson direction. In this case we have
%(\ref{Vmtree}), (\ref{Vm0}),
%(\ref{VmT})
\[
V(\varphi,Q,T)\sim T^{2}h^{2}\varphi^{2},\:\varphi\sim0.\]
Because the phase transition in the meson directions is first order, it is
accomplished through quantum tunneling processes and hence is much more
strongly suppressed than the classical phase transition in the quark
directions.

\begin{figure}
  \centerline{\includegraphics[width=\textwidth]{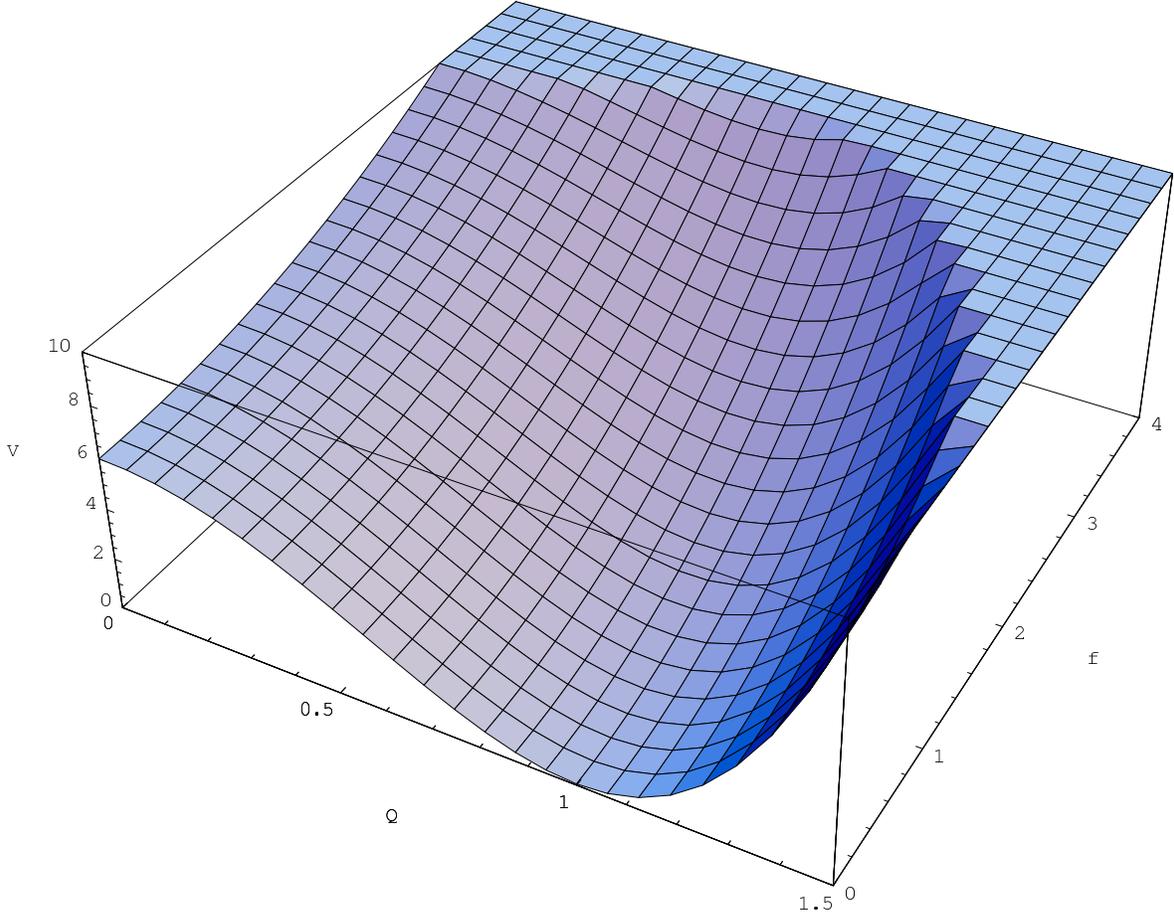}}
  \caption{\label{EffPot}Effective potential for every values of $\phi$ and $Q$ for $T\sim T_{c}^{Q}$}
\end{figure}

To gain a better understanding of the phase transition in the quarks direction 
we studied the finite temperature effective potential for every value of $\phi$
and $Q$ close to the critical temperature $T_{c}^{Q}$.
The result of this analysis is plotted in Fig. \ref{EffPot} From the shape of the effective 
potential around the origin we immediately realize that the flow of the vev happen
in the $Q$ direction and it is not possible for the vev to flow in the $\phi$ direction.

We are now in the position to form an idea about the phase history as
the universe
cools down. Let's suppose that we are starting at a temperature $T\gg
T_{c}$. We could for example be in the reheating phase after inflation.
At this temperature, the
origin
of field space is a minimum for the
finite temperature effective potential $V(\varphi,Q,T)$, figure \ref{T1}.
This is qualitatively plausible, since the massless fields make the
biggest contribution to entropy.
We also make the assumption that when we start off at this high
temperature,
the mesons $\varphi$ and the quarks $Q$ are localized around
the origin of field space: $\varphi=0, Q=0$ when $T\gg
T_{c}$
%\footnote{This
%assumption
%about the initial conditions is
%questionable, because the scale-invariance of the inflationary spectrum
%($\sim 1/k^3$) suggests that
%the scalar field values after inflation are not too sharply peaked around
%$k\sim 0$. But to have a complete understanding of the initial
%conditions, we need a better grasp of the actual dynamics of Seiberg
%transition and beyond, so we will stick with this simplistic
%assumption.}.
As the temperature decreases to $T=T_{c}^{Q}$,
the curvature of the effective potential $V(\varphi,Q,T)$ at the
origin becomes negative in the $Q$ direction but it stays positive
in the $\varphi$ direction:
\[
\left(\frac{\partial^{2}V}{\partial
Q^{2}}\right)_{\varphi=0,Q=0,T=T_{c}}<0,\]
\[
\left(\frac{\partial^{2}V}
{\partial\varphi^{2}}\right)_{\varphi=0,Q=0,T=T_{c}}>0.\]
Also, at $T=T_{c}^{Q}$, a new minimum $Q_{m}$ forms
in the $Q$ direction,
see figure \ref{T2}. As a consequence, a phase transition
occurs and the fields move to the newly formed minimum $Q_{m}$.
As the temperature of the universe continues to decrease, we eventually
arrive at $T\sim0$, see figure \ref{T3}, and the minimum $Q_{m}$ becomes
the (meta-stable) non-supersymmetric
vacuum $Q_{m}^{0}=\left(\begin{array}{c}
\mu\unit_{N}\\
0\end{array}\right)$. On the meson side, as the temperature drops, the
minimum $\varphi_{m}$
becomes the supersymmetric vacuum
$\varphi_{m}^{0}=\frac{\mu}{h}\frac{1}{\epsilon^{(N_{f}-3N)/(N_{f}-N)}}$,
but thankfully, phase-transition into the susy phase is suppressed
by tunneling at all stages. In writing the expression for the
susy-vacuum $\varphi_{m}^{0}$, we use the Intriligator et al.
convention, with $\epsilon\equiv\mu/\Lambda_m$ where $\Lambda_m$ is
the dynamically generated scale of our (infrared) theory. As mentioned
earlier, it is the
scale of the Landau pole.

Thus, the phase structure of the theory seems to imply that for
reasonably tame initial conditions for the scalar quarks and mesons
(namely, they start off near the origin of field space),
the phase transitions lead us into the susy-breaking vacuum at
$T\sim0$: \[
\varphi=0,\: Q_{m}^{0}=\left(\begin{array}{c}
\mu\unit_{N_{f}}\\
0\end{array}\right).\]

\section{Conclusions and Future Directions}

We found that the phase-structure of the free energy seems to imply
that as the universe cools, we end up in the susy-breaking vacuum
\footnote{When we were about to submit our paper, we became aware of an article that studies 
the same subject with similar conclusion \cite{Abel:2006cr}. }
. In
this final section, we point out some caveats and limitations
inherent in our study. One of these, is the assumption
of thermal equilibrium. By working with the free-energy, we are
ignoring the possibility that the fields could evolve and interact with
the heat bath. The dynamics of the fields could result in overshoot or
undershoot around the vacua. But it seems unlikely that these effects
will destroy our conclusion if the scalars are starting at the origin
of field space, at least if the evolution of the universe is close to
adiabatic. That brings us to a related issue, which is that we don't 
have a good idea about what are good initial conditions for the scalars, 
as was discussed in the introduction.

When calculating the effective potential, we focussed on just the meson
and squark directions. There is the possibility that we would discover new
valleys and slopes if we were to make a map of the potential for all
field values. One possibility is to do a perturbative analysis
around the quark/meson directions to see whether these directions are
indeed
troughs and not saddles at all temperatures.

Another issue which could affect our conclusions, even though it is
likely to be less important, is the question of the time scales
of first order versus second order transitions. It seems likely that
the original argument of Intriligator et al. suggesting that the
meta-stable vacuum can be made parametrically stable against
tunneling, should work (perhaps with a suitable modification) for our
case too.

To deal with non-equilibrium situations and initial conditions, one
could work in the context of the real-time formulation of finite
temperature field theory where we can consider both dynamics and
thermal effects. There,
the interaction with the heat bath would be encoded in a friction-like
term. Preliminary investigations in this direction seems to imply that
at least for some choice of scalar initial conditions, we end up in
the meta-stable vacuum. We leave a more thorough analysis of this
issue for future
work.

\section{Acknowledgments}

We thank Patrick J. Fox and Jay G. Wacker for useful conversations near the beginning
of this work.
We would like to thank Uday Varadarajan for giving a talk on the ISS
paper and thereby triggering this work, and also for discussions. 
The research of W.F., L.M. and V.K. was supported by NSF under grant PHY 0071512 and 0455649.
The work of C. K. is supported in part by
IISN - Belgium (convention 4.4505.86), by the Belgian National Lottery,
by the European Commission FP6 RTN programme MRTN-CT-2004-005104 in which
C. K. is associated with V. U. Brussel, and by the Belgian Federal
Science Policy Office through the Interuniversity Attraction Pole P5/27.
The work of M. T. is supported in part by CAPES.

%When we submitted our paper, we became aware of an article that studies 
%the same subject with similar conclusion \cite{Abel:2006cr}. 

% ==========================================================================
%
%%%%%%%%%%%%%%%%%%%%%%%%%%%%%%%%%%%%%%%%%%%%%%%%%%%%%%%%%%%%%%%%%%%%%%%%%%%%
%                      REFERENCES                            %
%%%%%%%%%%%%%%%%%%%%%%%%%%%%%%%%%%%%%%%%%%%%%%%%%%%%%%%%%%%%%%%%%%%%%%%%%%%%
\newpage
%\bibliography{metasusy}

\end{document}